\documentclass[english, a4paper]{article}
\usepackage[T1]{fontenc}
\usepackage{babel}
\usepackage{amsmath}
\usepackage{amssymb}
\usepackage{graphics}
\usepackage{epsfig}

\makeatletter
\@addtoreset{chapter}{part}
\makeatother

\newcommand\lsim{\mathrel{\rlap{\lower4pt\hbox{\hskip1pt$\sim$}}
    \raise1pt\hbox{$<$}}}
\newcommand\gsim{\mathrel{\rlap{\lower4pt\hbox{\hskip1pt$\sim$}}
    \raise1pt\hbox{$>$}}}

\newcommand\be{\begin{equation}}
\newcommand\bea{\begin{eqnarray} \nonumber }
\newcommand\ee{\end{equation}}
\newcommand\eea{\end{eqnarray}}

\begin{document}

\title{Two centuries of trend following}

\author{Y. Lemp\'eri\`ere, C. Deremble, P. Seager, M. Potters, J. P. Bouchaud \\
Capital Fund Management, 23 rue de l'Universit\'e, 75007 Paris, France}

\date{\today}
\maketitle

\begin{abstract}
We establish the existence of anomalous excess returns based on trend following strategies across four asset classes (commodities, 
currencies, stock indices, bonds) and over very long time scales. We use for our studies both futures time series, that exist since 1960, and spot time series
that allow us to go back to 1800 on commodities and indices. The overall t-stat of the excess returns is $\approx 5$ since 1960 and $\approx 10$ since 1800, after accounting for the overall upward drift of these markets. The effect is very stable, both across time and asset classes. It makes the existence of trends one of the most statistically significant anomalies in financial markets. When analyzing the trend following signal further, we find a clear saturation effect for large 
signals, suggesting that fundamentalist traders do not attempt to resist ``weak trends'', but step in when their own signal becomes strong enough. Finally, we
study the performance of trend following in the recent period. We find no sign of a statistical degradation of long trends, whereas shorter trends have significantly withered.
\end{abstract}

\section{Introduction}

Are markets efficient, in the sense that all public information is included in current prices? If this were so, price changes would be totally unpredictable in the sense that no systematic excess return based 
on public information can be achievable. After decades of euphoria in economics departments ({\it There is no other proposition in economics which has more solid empirical evidence supporting it than the Efficient Market Hypothesis}, as M. Jensen famously wrote in 1978), 
serious doubts were raised by behavioral economists who established a long series of pricing ``anomalies'' \cite{Schwert}. The most famous of these anomalies (and arguably the most difficult to sweep under the rug) is the so-called ``excess volatility puzzle'', unveiled by R. Shiller and others \cite{Shiller,LeRoy}. Strangely (or wisely?) the 2013 Nobel committee decided not to take sides, and declared that markets are indeed efficient (as claimed by laureate E. Fama), but that the theory actually 
makes ``little sense'', as argued by Shiller, who shared the same prize!\footnote{Together with a third scientist, L. Hansen, who had not directly taken part in the debate}. (See also \cite{Thaler,Black,Summers} for insightful papers on this debate.)

In the list of long-known anomalies, the existence of trends plays a special role. First, because trending is the exact 
opposite of the mechanisms that should ensure that markets are efficient, i.e. reversion forces that drive prices back to the purported fundamental value. Second, because persistent returns validate a dramatically 
simple strategy, {\it trend following}, which amounts to buying when the price goes up, and selling when it goes down. Simple as it may be \cite{turtle}, this strategy is at the heart of the activity of CTAs (Commodity Trading Advisors  \cite{trendCTA}), an industry that now manages (as of Q4, 2013) no less than 325 B\$, representing around $16 \%$ of the total amount of assets of the hedge fund industry, and accounting for several percent of the daily activity of futures markets.\footnote{Futures markets allow one to go short as easily as to go long. Therefore both up-trends and down-trends can equally be exploited.} \cite{CTA} These numbers are by no means small, and make it hard for efficient market enthusiasts to dismiss this 
anomaly as economically irrelevant.\footnote{M. Jensen (1978) actually stressed the importance of trading profitability in assessing market efficiency. In particular, if anomalous return behavior is not definitive enough for an efficient trader to make money trading on it, then it is not economically significant.} The strategy is furthermore deployed over a wide range of instruments (indices, bonds, commodities, currencies...) with positive reported performance over long periods, suggesting that the anomaly is to a large extent universal, both across epochs and asset classes.\footnote{Note that the excess return of trends cannot be classified as risk-premium either, see \cite{MoMR,RiskPremium}. On the contrary, trend following is correlated with ``long-vol'' strategies.} This reveals an extremely persistent, universal bias in the behaviour of investors who appear to hold ``extrapolative expectations'', 
as argued in many papers coming from different strands of the academic literature:
see e.g. \cite{deLong,Stein,Kirman,Smith,Hommes,Hirshleifer,Cont,Shleifer,Hirshleifer2} and refs. therein. 

Many academic studies have already investigated this trend anomaly on a wide range of assets, and have convincingly established its statistical significance in the last few decades \cite{trendCMD, trendRP}. Recently, this time horizon has been extended to 100 years in ref. \cite{trendAQR}, and the effect still exists unabated. The aim of the present paper is to extend the time horizon even further, to 200 years, as far in the
past as we have been able to go in terms of data. We find that the amplitude of the effect has been remarkably steady over two centuries. This also allows us to assess the recent weakening of the effect (as testified by the 
relatively poor performance of CTAs over the last five years). We show that the very recent past is fully compatible with a statistical fluctuation. Although we cannot exclude that this recent period is a precursor of the ``end of trends'', we argue theoretically that this is an unlikely scenario. We give several mechanisms that could explain the existence and persistence of these trends throughout history. 

Note that trends not only exist for market factors such as indices, bonds, currencies, etc., but also cross-sectionally in stock markets. The so-called ``momentum 
anomaly'' consists in buying the past winners and selling the past losers in a market neutral way, with again a high statistical significance across many decades and different geographical zones \cite{MoM1,MoM2,MoM3}, and \cite{MoMR} for a recent review. Although interesting in its own right (and vindicating the hypothesis that trend following is universal \cite{AQR}), we will not study this particular aspect of trend following in the present paper. 

The outline of the paper is as follows. In the next section, we define the trend following indicator used for this study, and test its statistical significance on available futures data. We start with futures since they are the preferred instruments of trend followers in finance. Also, their prices are unambiguously defined by transparent market trades, and not the result of a proprietary computation. 
In the following part, we carefully examine, for each asset class, how the available time-series can be extended as far in the past as possible. We then present our results over two centuries, 
and show how exceptionally stable long trends have been. We examine more in depth the linearity of the signal, and find that the trend predictability in fact saturates for large values of the signal, which is
needed for the long term stability of markets. We finally discuss the significance of the recent performance of the trend in light of this long-term simulation.

\section{Trend-following on futures since 1960}

\subsection{Measuring trends}

We choose to define our trend indicator in a way similar to simulating a constant risk trading strategy (without costs). More precisely, we first define the reference price level at time $t$, $\langle p \rangle_{n,t}$, as an exponential moving average of past prices (excluding $p(t)$ itself) with a decay rate equal to $n$ months. Long simulations can often only be performed on monthly data, so we use monthly closes. The signal $s_n(t)$ at the beginning of month $t$ is constructed as:
\be\label{signal}
s_n(t) = \frac{p(t-1) - \langle p \rangle_{n,t-1}}{\sigma_n(t-1)},
\ee
where the volatility $\sigma_n$ is equal to the exponential moving average of the absolute monthly price changes, with a decay rate equal to $n$ months.
The average strength of the trend is then measured as the statistical significance of 
fictitious profits and losses (P\&L) of a {\it risk managed} strategy that buys or sells (depending on the sign of $s_n$) a quantity $\pm \sigma_n^{-1}$ of the underlying contract $\alpha$:\footnote{We call this a fictitious P\&L since no attempt is made to model any realistic implementation costs of the 
strategy.}
\be \label{PandL}
Q_n^\alpha(t) = \sum_{t' < t} {\rm{sign}}[s_n(t')] \times \frac{p(t'+1) - p(t')}{\sigma_n(t'-1)}.
\ee
In the rest of the paper, we will focus on the choice $n=5$ months, although the dependence on $n$ will be discussed. Of course, different implementations can be proposed. However, 
the general conclusions are extremely robust against changes of the statistical test  or of the implemented strategy (see \cite{trendCMD, trendRP, trendCTA} for example).

In the following, we will define the Sharpe ratio of the P\&L as its average return divided by its volatility, both annualized. Since the P\&L does not include interest earned on the capital, and futures are self-financed instruments, we do not need to subtract the risk-free rate to compute the Sharpe ratio. The ``t-stat'' of the 
P\&L (i.e. the fact that the average return is significantly different from zero) is therefore given by the Sharpe ratio times $\sqrt{N}$, where $N$ is the number of years over which the strategy is active. 
We will also define the drift $\mu$ of a time series as the average daily return of the corresponding instrument, which would be the P\&L of the long-only strategy if financing costs were to be neglected.  

\subsection{The pool of assets}

Since we wish to prove that trend following is a universal effect not restricted to any one asset, we would like to test this signal on as large a pool as possible. This is also important in practice, since 
diversification plays an important role in the performance of CTAs. However, since the purpose of this paper is to back-test the trend on a very large history, we voluntarily limit ourselves to the contracts for which a long 
dataset is available. This naturally makes the inclusion of emerging markets more difficult. Therefore, for indices, bonds, and currencies, we only consider the following 7 countries: Australia, Canada,  Germany, Japan, Switzerland, United Kingdom and United States. We believe the results of this section would only be improved by the choice of a wider pool. 

We also need to select a pool of commodities. In order to have a well-balanced pool, we chose the following 7 representative contracts: Crude oil, Henry Hub Natural Gas, Corn, Wheat, Sugar, 
Live Cattle and Copper.

In summary, we have a pool made up of 7 commodity contracts, 7 10-year bond contracts, 7 stock index contracts and 6 currency contracts. All the data used in the current paper comes from GFD (Global Financial Data, {\tt
www.globalfinancialdata.com}).

\subsection{The results}

\begin{figure}
\begin{center}
\epsfig{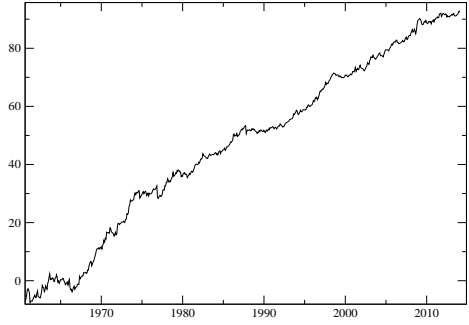}
\caption{Fictitious P\&L, as given by Eq. (\ref{PandL}) of a 5 month trend-following strategy on a diversified pool of futures. t-stat=5.9 (corresponding to a Sharpe ratio=0.8), de-biased t-stat=5.0}
\label{pnlALL}
\end{center}
\end{figure}

Our history of futures starts in 1960, mostly with commodities. As we can see from Fig. \ref{pnlALL}, the aggregated performance $\sum_\alpha Q_n^\alpha(t)$ looks well distributed in time, with an overall 
t-stat of 5.9, which is highly significant. The Sharpe ratio and t-stat are only weakly dependent on $n$, see Table \ref{SRsect}. 

However, one might argue that this comes from the trivial fact that there is an overall drift $\mu$ in most of these time series (for example, the stock market tends to
go up over time). It is therefore desirable to remove this ``long'' bias, by focussing on the {\it residual} of the trend following P\&L when the $\beta$ with the long-only 
strategy has been factored in. In fact, the correction is found to
be rather small, since the trend following P\&L and the long only strategy are only +15 \% correlated. Still, this correction decreases slightly the overall t-stat of the trend following performance to 5.0. 

In order to assess the significance of the above result, we break it down into different sectors and decades. As shown in Tables \ref{SRsect} and \ref{SRtime} the t-stat 
of the trend following strategy is above 2.1 for all sectors and all decades, and above 1.6 when de-biased from the drift $\mu$. Therefore, the performance shown in Fig. \ref{pnlALL} 
is well distributed across all sectors and periods, which strongly supports the claim that the existence of trends in financial markets is indeed universal.  One issue, though, is that our history of futures 
only goes back 50 years or so, and the first 10 years of those 50 is only made up of commodities. In order to test the stability and universality of the effect, 
it is desirable to extend the time series to go back further in the past, in order to span many economic cycles and different macro-environments. This is the goal of the next section, which provides a convincing 
confirmation of the results based on futures.

\begin{table}
\begin{center}
\begin{tabular}{|c|c|c|c|}
\hline
time-scale $n$ (months) & SR (T) & t-stat (T) & t-stat (T$^*$) \\
\hline
2 & 0.8 & 5.9 & 5.5 \\
3 & 0.83 & 6.1 & 5.5 \\
5 & 0.78 & 5.7 & 5.0 \\
7 & 0.8 & 5.9 & 5.0 \\
10 & 0.76 & 5.6 & 5.1 \\
15 & 0.65 & 4.8 & 4.5 \\
20 & 0.57 & 4.2 & 3.3 \\
\hline
\end{tabular}
\caption{Sharpe ratio and t-stat of the trend (T) and t-stat of the de-biased trend (T$^*$) for different time horizons $n$ (in months), since 1960.}
\label{SRscale}
\end{center}
\end{table}

\begin{table}
\begin{center}
\begin{tabular}{|c|c|c|c|c|c|c|}
\hline
sector & SR (T) & t-stat (T) & t-stat (T$^*$) & SR ($\mu$) & t-stat ($\mu$) & start date \\
\hline
Currencies & 0.57 & 3.6 & 3.4 & 0.05 & 0.32 & 05/1973 \\
Commodities & 0.8 & 5.9 & 5.0 & 0.33 & 2.45 & 01/1960 \\
Bonds & 0.49 & 2.8 & 1.6 & 0.58 & 3.3 & 05/1982 \\
Indices & 0.41 & 2.3 & 2.1 & 0.4 & 2.3 & 01/1982 \\
\hline
\end{tabular}
\caption{Sharpe ratio and t-stat of the trend (T) for $n=5$, of the de-biased trend (T$^*$) and of the drift component $\mu$ of the different
sectors, and starting date for each sector.}
\label{SRsect}
\end{center}
\end{table}

\begin{table}
\begin{center}
\begin{tabular}{|c|c|c|c|c|c|}
\hline
period & SR (T) & t-stat (T) & t-stat (T$^*$) & SR ($\mu$) & t-stat ($\mu$) \\
\hline
1960-1970 & 0.66 & 2.1 & 1.8 & 0.17 & 0.5 \\
1970-1980 & 1.15 & 3.64 & 2.5 & 0.78 & 2.5 \\
1980-1990 & 1.05 & 3.3 & 2.85 & -0.03 & -0.1 \\
1990-2000 & 1.12 & 3.5 & 3.03 & 0.79 & 2.5 \\
>2000 & 0.75 & 2.8 & 1.9 & 0.68 & 2.15 \\
\hline
\end{tabular}
\caption{Sharpe ratio and t-stat of the trend (T) for $n=5$, of the de-biased trend (T$^*$) and of the drift component $\mu$ for each decade.}
\label{SRtime}
\end{center}
\end{table}

\section{Extending the time series: a case-by-case approach}

We now try to find proxies for the futures time series that are reasonably correlated with the actual futures prices on the recent period but allow us to go back in the past a lot further. Natural candidates are spot prices on currencies, stock indices and commodities, and government rates for bonds. We shall examine each sector independently. 
Before doing so, however, we should mention other important restrictions on the use of the historical data. First, we expect trends to develop only on freely-traded instruments, where price evolution 
is not distorted by state interventions. Also, we require a certain amount of liquidity, in order to have meaningful prices. 
These two conditions, free-floating and liquid assets, will actually limit us when we look back in the distant past. 

\subsection{Currencies}

The futures time series goes back to 1973. In the previous period (1944-1971), the monetary system operated under the rules set out in the Bretton Woods agreements. According to these international treaties, 
the exchange rates were pegged to the USD (within a 1\% margin), which remained the only currency that was convertible into gold at a fixed rate. 
Therefore, no trend can be expected on these time series where prices are limited to a small band around a reference value.  

Prior to this, the dominant system was the Gold Standard. In this regime, the international value of a currency was determined by a fixed relationship with gold. 
Gold in turn was used to settle international accounts. In this regime as well, we cannot expect trends to develop, since the value of the currency is essentially fixed by its conversion rate with gold. 
In the 1930s, many countries dropped out of this system, massively devaluing in a desperate attempt to manage the consequences of the Great Depression (the ``beggar thy neighbor'' policy). 
This also led to massively managed currencies, with little hope of finding any genuine trending behavior. 

All in all, therefore, it seems unlikely that one can find a free-floating substitute for our futures time-series on foreign exchange, prior to 1973. 

\subsection{Government rates}

Government debt (and default!) has been around for centuries \cite{RR}, but in order to observe a trend on interest rates, one needs a liquid secondary market, on which the debt can be exchanged at all times. 
This is a highly non-trivial feature for this market. Indeed, throughout most of the available history, government debt has been used mostly as a way to finance extraordinary liabilities, such as wars. 
In other periods of history, debt levels gradually reduced, as the principals were repaid, or washed away by growth (as debt levels are quoted relative to GDP).

\begin{figure}
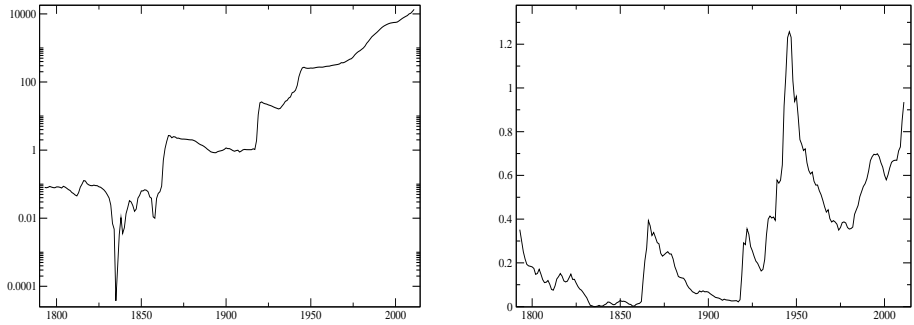

\begin{center}
\epsfig{file=USdebt.eps,height=0.2\textheight, width=0.45\textwidth,angle=0}\hskip 1cm
\epsfig{file=USdebt2.eps,height=0.2\textheight, width=0.45\textwidth,angle=0}
\caption{Global debt of the US government, in USD billions and as a fraction of GDP}
\label{USdebt}
\end{center}
\end{figure}

As a typical example, one can see in Fig. \ref{USdebt} that the US debt, inherited from the War of Independence, practically fell to 0 in 1835-36, during the Jackson presidency. 
There is another spike in 1860-65, during the American Civil War, which then gets gradually washed away by growth. We have to wait until the first World War to see a significant increase in debt which then persists until today. 
Apart from Australia, whose debt has grown at a roughly constant rate, and Japan, whose turning point is around 1905 and the war with Russia, the situation in all other countries is similar to that of the US. 
From this point onwards, the debt has never been repaid in its totality in any of the countries we consider in this study, and has mostly been rolled from one bond issuance to the next. 

Another more subtle point can explain the emergence of a stable debt market: at the beginning of the XXth century, the monetary policy (in its most straightforward sense: the power to print money) was separated from the executive instances and attributed to central banks, supposedly independent from the political power (see Fig. \ref{CBdates}). This move increased the confidence in the national debt of these countries, and helped boost subsequent debt levels. 

All of this leads us to the conclusion that the bond market before 1918 was not developed enough to be considered as ``freely traded and liquid''. Therefore, we start our interest rate time-series in 1918. 
We should note as well that we exclude from the time-series the war and immediate post-war period in Japan and Germany, where the economy was heavily managed, therefore leading to price distortions.

\begin{table}
\begin{center}
\begin{tabular}{|c|c|}
\hline
country & start \\
\hline
USA & 1913 \\
Australia & 1911 \\
Canada & 1935 \\
Germany & 1914 \\
Switzerland & 1907 \\
Japan & 1904 \\
United Kingdom & 1844 \\
\hline
\end{tabular}
\caption{Starting date of the central bank's monopoly on the issuance of notes. The bank of England does not have this monopoly in Scotland and Ireland, 
but regulates the commercial banks that share this privilege.}
\label{CBdates}
\end{center}
\end{table}

\subsection{Indices and commodities}

For these sectors, the situation is more straightforward. Stocks and commodities have been actively priced throughout the XIXth century, so it is relatively easy to get clean, well-defined prices. As we can see from Tables \ref{IDXstart} and \ref{CMDstart}, we can characterize trend following strategies for over more than 2 centuries on some of these time-series. Apart from some episodes that we excluded, like WWII in Germany or Japan where the stock market was closed, or the period through which the price of Crude was fixed (in the second half of the XXth century), the time-series are of reasonably good quality, i.e. 
prices are actually moving (no gaps) and there are no major outliers. 

\begin{table}
\begin{center}
\begin{tabular}{|c|c|}
\hline
country & start \\
\hline
USA & 1791 \\
Australia & 1875 \\
Canada & 1914 \\
Germany & 1870 \\
Switzerland & 1914 \\
Japan & 1914 \\
United Kingdom & 1693 \\
\hline
\end{tabular}
\caption{Starting date of the spot index monthly time-series for each country.}
\label{IDXstart}
\end{center}
\end{table}

\begin{table}
\begin{center}
\begin{tabular}{|c|c|}
\hline
commodity & start \\
\hline
Crude oil & 1859 \\
Natural Gas & 1986 \\
Corn & 1858 \\
Wheat & 1841 \\
Sugar & 1784 \\
Live Cattle & 1858 \\
Copper & 1800 \\
\hline
\end{tabular}
\caption{Starting date of the spot price for each commodity.}
\label{CMDstart}
\end{center}
\end{table}

\subsection{Validating the proxies}

We now want to check that the time series selected above, essentially based on spot data on 10 year government bonds, indices and commodities, yield results that are very similar to the ones we obtained with futures. 
This will validate our proxies and allow us to extend, in the following section, our simulations to the pre-1960 period.

\begin{figure}
\begin{center}
\epsfig{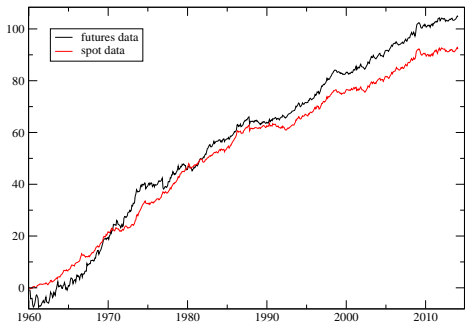}
\caption{Trend on spot and on futures prices. As one can see, the overall agreement since the late 60's (when the number of traded futures contracts becomes 
significant) is very good, although the average slope on spots is slightly smaller, as expected.}
\label{pnlSPOTFUT}
\end{center}
\end{figure}

In Fig. \ref{pnlSPOTFUT} we show a comparison of the trend applied to futures prices and to spot prices in the period of overlapping coverage between the two data sets. From 1982 onwards we have futures in all 4 sectors and the correlation is measured to be 91\%, which we consider to be acceptably high. We show the correlations per sector calculated since 1960 in Table \ref{corrSPOTFUT} and observe that the correlation remains high for indices and bonds but is lower for commodities with a correlation of 65\%. We know that the difference between the spot and futures prices is the so called ``cost of carry'' which is absent for the spot data, this additional term being especially significant and volatile for commodities. We find however that the level of correlation is sufficiently high to render the results meaningful. In any case the addition of the cost of carry can only improve the performance of the trend on futures and any conclusion regarding trends on spot data will be further confirmed by the use of futures data. 

We therefore feel justified in using the spot data to build statistics over a long history. We believe that the performance will be close to (and in any case, worse than) that on real futures, in particular because average financing costs are small, as illustrated by Fig. \ref{pnlSPOTFUT}.

\begin{table}
\begin{center}
\begin{tabular}{|c|c|}
\hline
sector & spot-future correlation\\
\hline
Commodities & 0.65 \\
Bonds & 0.91 \\
Indices & 0.92 \\
\hline
\end{tabular}
\caption{Correlation between spot and futures trend-following strategies. As we can see, even though the ``cost of carry'' plays an important role for commodities, the trends are still highly correlated.}
\label{corrSPOTFUT}
\end{center}
\end{table}

\section{Trend over two centuries}

\subsection{Results of the full simulation}

\begin{figure}
\begin{center}
\epsfig{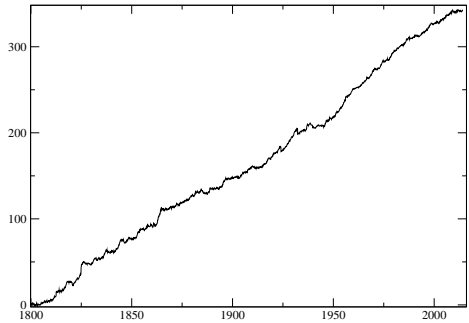}
\caption{Aggregate performance of the trend on all sectors. t-stat=10.5, de-biased t-stat=9.8, Sharpe ratio=0.72.}
\label{pnlALLlong}
\end{center}
\end{figure}

The performance of the trend-following strategy defined by Eq. (\ref{PandL}) over the entire time period (two centuries) is shown in Fig. \ref{pnlALLlong}. It is visually clear that the performance is  highly significant. This is confirmed by the value of the t-stat which is found to be above $10$, and $9.8$ when de-biased from the long only contribution, i.e. the t-stat of ``excess'' returns. For comparison, the t-stat of the drift $\mu$ of the same time series is $4.6$. As documented in Table \ref{SRall}, the performance is furthermore significant on each individual sector, with a t-stat of $2.9$ or higher, and $\geq 2.7$ when the long bias is removed. Note that the de-biased t-stat of the trend is in fact higher than the 
t-stat of the long-only strategy, with the exception of commodities, where it is slightly worse ($3.1$ vs. $4.5$).

The performance is also remarkably constant over two centuries: this is obvious from Fig. \ref{pnlALLlong}, and we report the t-stat for different periods in
Table \ref{SR50}. The overall performance is in fact positive over every decade in the sample, see Fig. \ref{10Yroll} below. The increase in performance in the second half of the simulation probably comes from the fact that we have more and more products as time goes on (indeed, government yields and currencies both start well into the XXth century).

\begin{table}
\begin{center}
\begin{tabular}{|c|c|c|c|c|c|c|}
\hline
sector & SR (T) & t-stat (T) & t-stat (T$^*$) & SR ($\mu$) & t-stat ($\mu$) & start date\\
\hline
Currencies & 0.47 & 2.9 & 2.9 & 0.1 & 0.63 & 1973 \\
Commodities & 0.28 & 4.1 & 3.1 & 0.3 & 4.5 & 1800 \\
Bonds & 0.4 & 3.9 & 2.7 & -0.1 & -1 & 1918 \\
Indices & 0.7 & 10.2 & 6.3 & 0.4 & 5.7 & 1800 \\
\hline
\end{tabular}
\caption{Sharpe ratio and t-stat of the trend (T), of the de-biased trend (T$^*$) and of the drift component $\mu$ of the different
sectors. We also list the starting date for each sector}
\label{SRall}
\end{center}
\end{table}

\begin{table}
\begin{center}
\begin{tabular}{|c|c|c|c|c|}
\hline
period & SR (T) & t-stat (T) & SR ($\mu$) & t-stat ($\mu$)  \\
\hline
1800-1850 & 0.6 & 4.2 & 0.06 & 0.4 \\
1850-1900 & 0.57 & 3.7 & 0.43 & 3.0  \\
1900-1950 & 0.81 & 5.7 & 0.34 & 2.4 \\
>1950 & 0.99 & 7.9 & 0.41 & 2.9 \\
\hline
\end{tabular}
\caption{Sharpe ratio and t-stat of the trend and of the drift $\mu$ over periods of 50 years.}
\label{SR50}
\end{center}
\end{table}

\subsection{A closer look at the signal}

\begin{figure}
\begin{center}
\epsfig{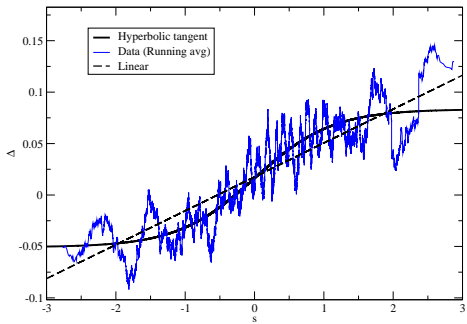}
\caption{Fit of the scatter plot of $\Delta(t)=p(t+1)-p(t)$ as a function of $s_n(t)$, for $n=5$ months, and for futures data only. We do not show the 240,000 points on which the fits are performed, but rather a running average over 
5000 consecutive points along the $x$-axis. We also show 
the results of a linear and hyperbolic tangent fit. Note the positive intercept $a \approx 0.02$ that indicates the overall positive long-only bias. The best fit to the data is provided by the 
hyperbolic tangent, which suggests a saturation of the signal for large values. }
\label{scatter}
\end{center}
\end{figure}

It is interesting to delve deeper into the predictability of the trend following signal $s_n(t)$, defined in Eq. (\ref{signal}). Instead of 
computing the P\&L given by Eq. (\ref{PandL}), one can instead look at the scatter plot of $\Delta(t)=p(t+1)-p(t)$ as a function of $s_n(t)$. This 
gives a noisy blob of points with, to the naked eye, very little structure. However, a regression line through the points lead to a statistically
significant slope, i.e. $\Delta(t) = a + b s_n(t) + \xi(t)$, where $a=0.018 \pm 0.003$, $b=0.038 \pm 0.002$ and $\xi$ is a noise term. The fact that $a > 0$ is equivalent to 
saying that the long only strategy is, on average, profitable, whereas $b > 0$ indicates the presence of trends. 
However, it is not {\it a priori} obvious that one should expect a linear relation between $\Delta$ and $s_n$. Trying a cubic regression gives a very small coefficient for the $s_n^2$ term 
and a clearly negative coefficient for the $s_n^3$ term, indicating that strong signals tend to flatten, as suggested by a running average of the signal shown in Fig. \ref{scatter}. 
However, the strong {\it mean-reversion} that such a negative cubic contribution would predict for large $s_n$'s is suspicious. We have therefore rather tried to model a non-linear saturation through a 
hyperbolic tangent (Fig. \ref{scatter}):
\be
\Delta(t) = a + bs^* \tanh \left(\frac{s_n(t)}{s^*}\right) + \xi(t), 
\ee
which recovers the linear regime when $|s_n| \ll s^*$ but saturates for $|s_n| > s^*$. This non-linear fit is found to be better than the cubic fit as well 
as the linear fit as it prefers a finite value $s^* \approx 0.89$ and now $b \approx 0.075$ (a linear fit is recovered in the limit $s^* \to \infty$). 
Interestingly, the value of $a,b$ and $s^*$ hardly change when $n$ increases from $2.5$ months to $10$ months.

\subsection{A closer look at the recent performance}

\begin{figure}
\begin{center}
\epsfig{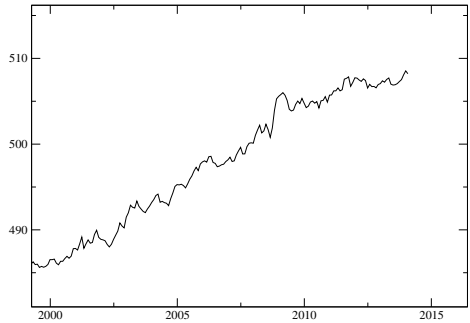}
\caption{Recent performance of the trend. Since 2011, the strategy is virtually flat.}
\label{pnlTRENDrecent}
\end{center}
\end{figure}

The plateau in the performance of the trend over the last few years (see Fig. \ref{pnlTRENDrecent}) has received a lot of attention from CTA managers and investors. Amongst other explanations, the over-crowdedness of the strategy has been frequently evoked to explain this relatively poor performance. We now want to reconsider these conclusions in the context of the long term simulation. 

First, it should not come as a surprise that a strategy with a historical Sharpe ratio below 0.8 shows relatively long drawdowns. In fact, the {\it typical}
duration of a drawdown is given by $1/S^2$ (in years) for a strategy of Sharpe ratio $S$. This means that for a Sharpe of $0.7$, typical drawdowns last two years while
drawdowns of 4 years are not exceptional (see \cite{BP,ddCFM} for more on this topic). 

\begin{figure}
\begin{center}
\epsfig{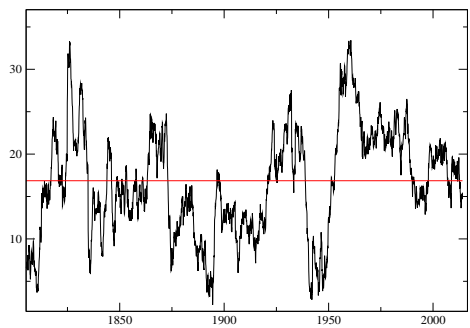}
\caption{10-year cumulated performance of the trend (arbitrary units). The horizontal line is the historical average.}
\label{10Yroll}
\end{center}
\end{figure}

To see how significant the recent performance is, we have plotted in Fig. \ref{10Yroll} the average P\&L between time $t - 10Y$ and time $t$. 
We find that, though we are currently slightly below the historical average, this is by no means an exceptional situation. A much worse performance was 
in fact observed in the 1940s. Fig. \ref{10Yroll} also reveals that the 10-year performance of trend-following has, as noted above, never been negative in two centuries, which is again 
a strong indication that trend following is ingrained in the evolution of prices. 

\begin{figure}
\begin{center}
\epsfig{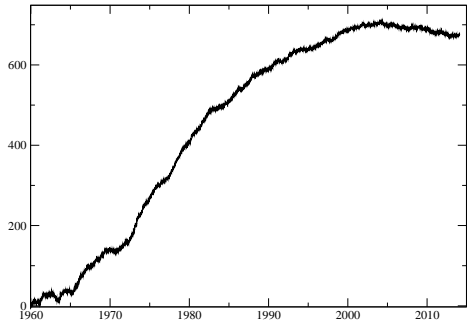}
\caption{Performance of a 3-day trend on futures contracts since 1970. The effect seems to have completely disappeared since 2003 (or has maybe even
inverted).}
\label{3days}
\end{center}
\end{figure}

The above conclusion is however only valid for long-term trends, with a horizon of several months. Much shorter trends (say over three days) have significantly decayed since 1990, see 
Fig. \ref{3days}. We will now propose a tentative interpretation of these observations.

\subsection{Interpretation}

The above results show that long term trends exist across all asset classes and are stable in time. As mentioned in the introduction, trending behaviour is also observed in the idiosyncratic component of individual stocks \cite{MoM1,MoM2,MoM3,MoMR}. What can explain such universal, persistent behaviour of prices? One can find two (possibly complementary) broad families of interpretation in the literature. One explanation assumes that agents under-react to news, and only progressively include
the available information into prices \cite{Stein,Hirshleifer}. An example of this could be an announced sequence of rate increases by a central bank over several months, which is not immediately reflected in bond prices because market participants tend to only believe in what they see, and are slow to change their previous expectations (``conservatism bias''). In general, changes of policy (for governments, central banks or indeed companies) are slow and progressive. If correctly anticipated, prices should immediately reflect the end point of the policy change. Otherwise, prices will progressively follow the announced changes and this inertia leads to trends. 

Another, distinct mechanism is that market participants' expectations are directly influenced by past trends:
positive returns make them optimistic about future prices, and vice-versa. These ``extrapolative expectations'' are supported by ``learning to predict`'' experiments in artificial markets \cite{Smith,Hommes}, which show that linear extrapolation is a strongly anchoring strategy. In a complex world where information is 
difficult to decipher, trend following -- together with herding -- is one of the ``fast and frugal'' heuristics \cite{Gigerenzer} that most people are tempted to use \cite{JStatPhys}. Survey data also strongly point in this direction \cite{Shiller2,Shleifer,Menkhoff}.\footnote{Anecdotally, based on a long history of CFM inflows and outflows, our experience strongly suggests that professional investors have a strong tendency to ``chase performance'', i.e. to invest in CFM's funds after a positive rally and redeem after negative performance.} Studies of agent based models in fact show that the imbalance between trend following and fundamental pricing is crucial to account for some of the stylized facts of financial markets, such as fat tails and volatility clustering (see e.g. \cite{Lux,GB,Hommes2} and more recently \cite{Shleifer2}). Clearly, the perception of trends can lead to positive feedback trading which reinforces the existence of trends, rather than making them disappear \cite{deLong,Cont,WB}. 

On this last point, we note that the existence of trends {\it far predates} the explosion of assets managed by CTAs. The data shown above suggests that CTAs  
have neither substantially increased, nor substantially reduced, the strength of long term trends in major financial markets. While the degradation in recent performance, 
although not statistically significant, might be attributed to over-crowding of trending strategies, it is not entirely clear how this would happen in the 
``extrapolative expectations'' scenario which tends to be self-reinforcing (see e.g. \cite{WB} for an explicit model). If, on the other hand, under-reaction is
the main driver of trends in financial markets, one may indeed see trends disappear as market participants better anticipate long term policy changes (or indeed policy makers become more easily predictable).
Still, the empirical evidence supporting a behavioral trend following propensity seems to us strong enough to advocate extrapolative expectations over under-reactions. It would be interesting to build a detailed behavioral model that explains why the trending signal saturates at high values, as evidenced in Fig. \ref{scatter}. One plausible interpretation is that when prices become more obviously out of line, fundamentalist traders start stepping in, and this mitigates the impact of trend followers which are still lured in by the strong trend (see \cite{Cont,Lux,Shleifer} for similar stories).

\section{Conclusions}

In this study, we have established the existence of anomalous excess returns based on trend following strategies across all asset classes and over very long time scales. We first studied futures, as is customary, then spot data that allows us to go far back in history. We carefully justified our procedure, in particular by comparing the results on spot data in the recent period, which shows a strong correlation with futures, with very similar drifts. The only sector where we found no way to extend the history is for foreign exchange, since the idea of a free-floating currency is a rather recent one. 

We have found that the trend has been a very persistent feature of all the financial markets we have looked at. The overall t-stat of the excess returns is around 10 since 1800, after accounting for the long-only bias. Furthermore, the excess returns associated to trends cannot be associated to any sort of risk-premium \cite{MoMR,RiskPremium}. The effect is very stable, both across time and asset classes. It makes the existence of trends one of the most statistically significant anomalies in financial markets. When analyzing the trend following signal further, we found a clear saturation effect for large  signals, suggesting that fundamentalist traders do not attempt to resist ``weak trends'', but might step in when their own signal becomes strong enough.

We investigated the statistical significance of the recent mediocre performance of the trend, and found that this was actually in line with a long historical back-test. Therefore, the suggestion that long-term trend-following has become over-crowded is not borne out by our analysis, and is compatible with our estimate that CTAs only contribute a few percent of market volumes.  Still, the understanding of the behavioral causes of trends, and in particular the relative role of ``extrapolative expectations'' vs. ``under-reaction'' or ``conservative biases'' would allow one to form an educated opinion on the long term viability of trend-following strategies. It is actually not obvious how crowdedness would deteriorate trend-following strategies, since more trend-following should speed-up trends as managers attempt to ``front-run'' the competition. Fig. \ref{3days}, however, adds to the conundrum by showing that faster trends have actually progressively disappeared in recent years, without ever showing a period where they strengthened. Coming up with a plausible mechanism that explains how these fast trends have disappeared would be highly valuable in understanding the fate of trends in financial markets.

\section*{Acknowledgements} This work is the result of many years of research at CFM. Many colleagues must be thanked for their insights, in particular:  P. Aliferis, N. Bercot, A. Berd, D. Challet,  L. Dao, B. Durin, P. Horvai, L. Laloux, A. Landier, A. Matacz, D. Thesmar, T. Tu, and M. Wyart.


\begin{thebibliography}{10}

\bibitem{Schwert} G. W. Schwert, {\it Anomalies and market efficiency}, Handbook of the Economics of Finance
Volume 1, Part B, 939-974 (2003)

\bibitem{Shiller} R. J. Shiller, {\it Do stock prices move too much to be justified by subsequent
changes in dividends?}, American Economic Review 71, 421-436 (1981)

\bibitem{LeRoy} S.F. Leroy, R.D. Porter, {\it The present value
relation: Tests based on implied variance bounds}, Econometrica 49, 555 (1981)

\bibitem{Thaler} W. de Bondt, R. H. Thaler, {\it Does the stock market overreact?}, Journal
of Finance 42, 557-581 (1985)

\bibitem{Black} F. Black, {\it Noise}, Journal of Finance, 41 529-543 (1986).

\bibitem{Summers} L. Summers, {\it Does the Stock market rationally reflect fundamental values?}
Journal of Finance, XLI, (1986) 591

\bibitem{turtle} Michael Covel, \emph{The complete Turtle Trader}, HarperCollins, 2009.

\bibitem{trendCTA}
A.-N. Bartas, R. Kosowski, \emph{Momentum Strategies in Futures Markets and Trend-following Funds}, 
Paris December 2012 Finance Meeting EUROFIDAI-AFFI Paper, 
Available at SSRN: http://ssrn.com/abstract=1968996 or http://dx.doi.org/10.2139/ssrn.1968996.

\bibitem{CTA} M. Mundt, {\it Estimating the capacity of the managed futures industry}, in CTA Intelligence, {\bf 12}, 30 (March 2014) 


\bibitem{MoMR} J. Narasimhan, S. Titman, {\it Momentum} (2011).
Available at SSRN: http://ssrn.com/abstract=1919226

\bibitem{RiskPremium} Y. Lemp\'eri\`ere et al., {\it What is Risk Premium and how does it differ from alpha strategies?}, in preparation (2014)

\bibitem{deLong} J. DeLong, A. Bradford, A. Shleifer, L. H. Summers, and R. J. Waldmann, {\it Positive feedback investment
strategies and destabilizing rational speculation}, Journal of Finance 45:379-95 (1990).

\bibitem{Stein} H. Hong, J. Stein, {\it A Unified Theory of Underreaction, Momentum Trading, and Overreaction in Asset Markets}, The Journal of Finance
54, 2143-2184 (1999).

\bibitem{Kirman} A. Kirman, {\it Epidemics of opinion and speculative bubbles in financial markets},
in Mark P. Taylor, ed.: Money and Financial Markets (MacMillan) 1991; {\it Ants, rationality and recruitment}, 
Quarterly Journal of Economics 108, 137-156 (1993).

\bibitem{Smith} V. L. Smith, G. L. Suchanek, A. W. Williams, {\it Bubbles, crashes
and endogenous expectations in experimental spot asset markets}, Econometrica 56,
1119-51 1988

\bibitem{Hommes} 
C. Hommes, J. Sonnemans, J. Tuinstra, H. van de Velden, {\it  Expectations and bubbles in asset pricing
experiments}, J. Econ. Behav. Organ. 67, 116 (2008)


\bibitem{Hirshleifer} D. Kent, D. Hirshleifer, A. Subrahmanyam, \emph{Investor Psychology and Security Market Under- and Overreations}, Journal of Finance, 
53, 1839-85 (1998).

\bibitem{Cont} J.-P. Bouchaud, R. Cont, {\it A Langevin approach to stock market fluctuations and crashes}, Eur. J. Phys. B 6, 543 (1998)


\bibitem{Shleifer} R. Greenwood, A. Shleifer, {\it Expectations of Returns and Expected Returns}, 
Rev Fin Studies, to appear (2014)

\bibitem{Hirshleifer2} D. Hirshleifer, J. Yu, {\it Asset pricing in production economies with extrapolative expectations}, Working
Paper (2012)


\bibitem{trendCMD} A. C. Szakmary, Qian Shen, S. C. Sharma, 
\emph{Trend-Following Strategies in Commodity Futures: a Re-Examination}, 
Journal of Banking and Finance, 
Vol. 34-2, 409-426 (2010).

\bibitem{trendRP}
A. Clare, J. Seaton, P. N. Smith, S. Thomas, 
\emph{Trend Following, Risk Parity and Momentum in Commodity Futures}, 
2012,
Available at SSRN: http://ssrn.com/abstract=2126813 or http://dx.doi.org/10.2139/ssrn.2126813.


\bibitem{trendAQR}
B. Hurst, Y. H. Ooi, L. H. Pedersen, 
\emph{A century of Evidence on Trend-Following Investing}, AQR Working paper, 2012.

\bibitem{MoM1} J. Narasimhan, S. Titman, {\it Returns to buying winners and selling losers: Implications
for stock market efficiency}, Journal of Finance 48, 65-91 (1993).

\bibitem{MoM2} D. Kent, T. J. Moskowitz, {\it Momentum Crashes} Swiss Finance Institute Research Paper Series 13-61

\bibitem{MoM3} P. Barroso, P. Santa-Clara, {\it Momentum has its moments}, Working paper (2013)

\bibitem{AQR} C. S. Asness, T. J. Moskowitz, and L. H. Pedersen,  {\it Value and momentum
everywhere}, The Journal of Finance 58, 929-895 (2013).

\bibitem{RR}
C. M. Reinhart, K. S. Rogoff, 
\emph{This Time is Different: Eight Centuries of Financial Folly}, 
Princeton University Press,  2009.

\bibitem{BP} J. P. Bouchaud, M. Potters, {\it Theory of Financial Risk and Derivative Pricing}, Cambridge University Press (2003).

\bibitem{ddCFM} P. Seager, et al. \emph{The statistics of drawdowns}, in preparation, 2014.

\bibitem{Gigerenzer} G. Gigerenzer, D. Goldstein, {\it Reasoning the fast and frugal way: models of bounded rationality}. Psychol.
Rev. 103, 650 (1996)

\bibitem{JStatPhys} J. P Bouchaud, {\it Crises and collective socio-economic phenomena: simple models and challenges}, 
J. Stat. Phys. 151 567 (2013).

\bibitem{Shiller2} R. J. Shiller, {\it  Measuring bubble expectations and investor confidence}, Journal of Psychology and Financial
Markets 1:49-60 (2000)

\bibitem{Menkhoff} L. Menkhoff, {\it Are Momentum Traders Different? Implications for the Momentum Puzzle},
Discussion Paper No.448, May 2010


\bibitem{Lux} 
T. Lux, M. Marchesi, {\it  Volatility clustering in financial markets: a micro-simulation of interacting
agents}, Int. J. Theor. Appl. Finance 3, 675 (2000)

\bibitem{GB} I. Giardina, J.-P. Bouchaud, {\it Bubbles, crashes and intermittency in agent based market models}, Eur.
Phys. J. B 31, 421 (2003). Within their model, these authors show that without an element of trend-following, markets quickly 
reach an ``efficient'' stationary state where nothing much happens.

\bibitem{Hommes2} 
C. H. Hommes, {\it Heterogeneous agent models in economics and finance}, in Kenneth
L. Judd, and Leigh Tesfatsion, ed.: Handbook of Computational Economics (North-
Holland) Vol. 2: Agent-Based Computational Economics, 2006.

\bibitem{Shleifer2} N. Barberis, R. Greenwood, L. Jin, A. Shleifer. {\it X-CAPM: An extrapolative capital asset pricing model},
Working Paper (2013).

\bibitem{WB} 
M. Wyart, J.-P. Bouchaud, {\it Self-referential behaviour, overreaction and conventions in financial markets}
J. Econ. Behav. Organ. 63, 1 (2007)


\end{thebibliography}
\end{document}